\documentclass{ws-procs10x7}
\usepackage{balance}
\usepackage{bbm}
\usepackage{pifont}
\newcommand{\mnu}{\mathcal{M}_\nu}
\newcommand{\zx}{\mathbbm{Z}_2^{(\mathrm{aux})}}
\newcommand{\dsol}{\Delta m^2_\odot}
\newcommand{\datm}{\Delta m^2_\mathrm{atm}}
\newcolumntype{d}[1]{D{.}{.}{#1}}

\def\Journal#1#2#3#4{{\it #1} {\bf #2}, #3 (#4)}

\makeindex
\begin{document}

\title{Realizations of $\mu$--$\tau$ interchange symmetry}

\author{W. Grimus$^*$}

\address{Institute for Theoretical Physics, University of Vienna \\
Boltzmanngasse 5, A-1090 Vienna, Austria \\
$^*$E-mail: walter.grimus@univie.ac.at}

\twocolumn[\maketitle\abstract{
Some models for the lepton sector, based on seesaw extensions of the Standard
Model, are discussed in which the $\mu$--$\tau$ interchange symmetry 
is realized in various ways. The symmetries defining such models and their
characteristic predictions for lepton mixing are presented.
}
]

\section{Mass matrices with $\mu$--$\tau$ interchange symmetry}
At present, all data from neutrino oscillation measurements are
compatible with the lepton mixing angles\cite{tortola}
\begin{equation}\label{a}
\theta_{23} = 45^\circ 
\quad \mbox{and} \quad
\theta_{13} = 0^\circ.
\end{equation}
A neutrino mass matrix with these properties,
in the basis in which the charged-lepton mass
matrix $M_\ell$ is diagonal,
exhibits a $\mu$--$\tau$ (interchange)
symmetry---for early papers on this issue see
Refs.~\refcite{early,Z2model}. Assuming that the 
lepton flavours $\alpha$ are
ordered in the usual way with $\alpha = e,\, \mu,\, \tau$, the
$\mu$--$\tau$ symmetry for a matrix $M^{(\mathrm{S})}$ is 
formulated as\cite{ustron}
\begin{equation}\label{S}
T M^{(\mathrm{S})} T = M^{(\mathrm{S})} \Rightarrow 
M^{(S)} = \left( \begin{array}{ccc} x & y & y \\ y & z & w \\
y & w & z \end{array} \right),
\end{equation}
where $T$ is a permutation matrix, performing the flavour exchange 
$\mu \leftrightarrow \tau$.
If $M^{(\mathrm{S})}$ is conceived as a Majorana
neutrino mass matrix $\mnu$, it is easy to see that it 
predicts the mixing angles of Eq.~(\ref{a}). There are no further
predictions of $M^{(\mathrm{S})}$.\cite{Z2model}

With 
$\mu$--$\tau$ antisymmetry\cite{aizawa,anti}, one obtains
\begin{eqnarray}
&& \nonumber
T M^{(\mathrm{AS})} T = -M^{(\mathrm{AS})} \Rightarrow \\
&& \label{AS}
M^{(\mathrm{AS})} = \left( \begin{array}{ccc} 0 & p & -p \\ p & q & 0 \\
-p & 0 & -q \end{array} \right).
\end{eqnarray}
Assuming $\mnu = M^{(\mathrm{AS})}$, this 
matrix gives\cite{anti}
\begin{equation}\label{pred-anti}
\theta_{12} = \theta_{23} = 45^\circ, \;
m_1 = m_2,\; m_3 = 0.
\end{equation}
We gather from this result that $M^{(\mathrm{AS})}$ is not
suitable as a neutrino mass matrix, however, its predictions are not
excessively far from reality, if we assume $\theta_{13}$ to be
sufficiently small.

\section{$\mnu$ versus $\mnu^{-1}$}
\begin{enumerate}
\item\label{i}
Denoting the diagonalizing matrix of $\mnu$ by $U$ and assuming
$\det \mnu \neq 0$, we observe the relationship
\begin{eqnarray}
&& \nonumber 
U^T \mnu\, U = \hat m \equiv \mbox{diag}\,( m_1, m_2, m_3 )
\; \Leftrightarrow \\
&&
U^\dagger \left(\mnu \right)^{-1} U^* = \left( \hat m \right)^{-1}.
\end{eqnarray}
\item\label{ii}
Next, we assume the validity of the seesaw mechanism: 
$\mnu = -M_D^T M_R^{-1} M_D$ with the neutrino Dirac-mass matrix $M_D$
and the mass matrix $M_R$ of the right-handed neutrino singlet fields
$\nu_R$  whose mass Lagrangian is given by
\begin{equation}\label{LM}
\mathcal{L}_M(\nu_R) = \frac{1}{2}\, \nu_R^T C^{-1} M_R^* \nu_R + 
\mbox{H.c.} 
\end{equation}
If $M_\ell$ is diagonal and 
$M_D$ has the form $M_D = \mbox{diag}\,(a,b,b)$,
then $\mu$--$\tau$ symmetry (antisymmetry) of 
$\left(\mnu \right)^{-1}$ is equivalent to
$\mu$--$\tau$ symmetry (antisymmetry) of $M_R$.
\end{enumerate}
To impose $\mu$--$\tau$ symmetry on $M_R$ we simply have to require
invariance of $\mathcal{L}_M(\nu_R)$ under 
$\nu_R \to T  \nu_R$; for $\mu$--$\tau$ antisymmetry
the transformation is $\nu_R \to iT \nu_R$. 

With the assumptions of Item~(\ref{ii}) it makes sense to impose
conditions on the inverse mass matrix\cite{Z2model,lavoura} instead 
of on the mass matrix itself. 
Obviously, $\left( \mnu \right)^{-1}$ can be decomposed as
\begin{equation}\label{decomposition}
\left( \mnu \right)^{-1} = M^{(\mathrm{S})} + M^{(\mathrm{AS})}.
\end{equation}
This rather trivial observation is the point of departure for the
following discussion where we will present several models which
realize the $\mu$--$\tau$ interchange symmetry within seesaw extensions of the
Standard Model (SM). 

\section{The framework}
\label{framework}
We consider the lepton sector of the SM, enlarge the
scalar sector to three Higgs doublets $\phi_j$ ($j = 1,\,2,\,3$) and add three
right-handed neutrino singlets 
$\nu_{\alpha R}$ for the purpose of the seesaw
mechanism. The left-handed lepton doublets are denoted by 
$D_{\alpha L}$ and the right-handed charged-lepton singlets by
$\ell_{\alpha R} \equiv \alpha_R$.
We impose the following symmetries:
\begin{enumerate}
\item\label{fs}
The groups 
$U(1)_{L_\alpha}$ ($\alpha = e,\mu,\tau$) associated with the family
lepton numbers $L_\alpha$, 
or, alternatively, 
$D_L \to \mbox{diag}\, (1,\omega,\omega^2) D_L$ and, analogously, for
$\ell_R$ and $\nu_R$, with
$\omega = e^{2\pi i/3}$, corresponding to the symmetry group
$\mathbbm{Z}_3$.
\item\label{mu-tau}
The symmetry transformation
$D_L \to i^k T D_L$, $\ell_R \to i^k T \ell_R$, 
$\nu_R \to i^k T \nu_R$, $\phi_3 \to -\phi_3$, 
which either corresponds to the $\mu$--$\tau$ symmetry for $k=0$ or to the 
$\mu$--$\tau$ antisymmetry for $k=1$.
\item\label{aux}
An auxiliary symmetry $\zx$ defined by the
sign change of the fields 
$\nu_{\alpha R}$ ($\alpha = e,\, \mu,\, \tau$), $\phi_1$, $e_R$.
\end{enumerate}
It is easy to check that the most general Yukawa Lagrangian compatible
with these symmetries is given by
\begin{eqnarray}
\lefteqn{\mathcal{L}_Y(\phi) =} 
\nonumber \\ &&
- y_1 \bar D_{eL} \nu_{eR} \tilde\phi_1  
- y_2 \left( \bar D_{\mu L} \nu_{\mu R} + \bar D_{\tau L} \nu_{\tau R} \right)
\tilde\phi_1 
\nonumber \\ && 
- y_3 \bar D_{eL} e_R \phi_1
- y_4 \left( \bar D_{\mu L} \mu_R + \bar D_{\tau L} \tau_R \right) \phi_2
\nonumber \\ && \label{LY}
- y_5 \left( \bar D_{\mu L} \mu_R - \bar D_{\tau L} \tau_R \right) \phi_3
+ \mbox{H.c.}
\end{eqnarray}
Note that the symmetries of Item~(\ref{fs}) enforce diagonal Yukawa
couplings, Item~(\ref{mu-tau}) provides the $\mu$--$\tau$-symmetric
strucure of the couplings, and Item~(\ref{aux}) makes sure that 
$\phi_3$ does not couple to $\nu_R$; the latter point is important for
supplying the $\mu$--$\tau$-symmetric form $M_D = \mbox{diag}\,(a,b,b)$
of the neutrino Dirac-mass matrix. 

The $\mu$--$\tau$ (anti)symmetry is spontaneously broken by the VEV
of $\phi_3$, which allows for $m_\mu \neq m_\tau$.

\section{A model based on $S_3 \times \zx$}
The model presented in this section\cite{GL05} is based on the group 
$S_3$. We have the following representations: 
$D_L,\, \ell_R,\, \nu_R \in \underline{1} \oplus \underline{2}$,
$\phi_{1,2} \in \underline{1}$, $\phi_3 \in\underline{1}'$. We add a
complex scalar $\chi$ such that $(\chi, \chi^*) \in \underline{2}$. 
The connexion of $S_3$ with the symmetries of the previous section is
obtained via 
\begin{equation}
\underline{2}: \left\{
\begin{array}{l}
(\mathit{12}) \to 
\left( \begin{array}{cc} 0 & 1 \\ 1 & 0 \end{array} \right), \\
(\mathit{123}) \to 
\left( \begin{array}{cc} \omega & 0 \\ 0 & \omega^2 \end{array}
\right),
\end{array}
\right.
\end{equation}
where $(\mathit{12}),\, (\mathit{123}) \in S_3$. 
The cyclic permutation represents the $\mathbbm{Z}_3$ symmetry of
Item~(\ref{fs}), whereas 
the transposition $(\mathit{12})$ is mapped into the
$\mu$--$\tau$ symmetry of Item~(\ref{mu-tau}). The trivial
one-dimensional representation is denoted by 
$\underline{1}$ and 
$\underline{1}'$ is given by 
$(\mathit{12}) \to -1$, $(\mathit{123}) \to 1$.

Apart from the Lagrangian~(\ref{LY}), the symmetries allow Yukawa
couplings of the singlet scalar, described by the Lagrangian
\begin{eqnarray}
\lefteqn{\mathcal{L}_Y(\chi) = } \nonumber \\ 
&& 
y_\chi^\ast\, \nu_{eR}^T C^{-1}
\left( \nu_{\mu R} \chi^\ast + \nu_{\tau R} \chi \right) +
\\ &&
\frac{1}{2}\,z_\chi^\ast \left( \nu_{\mu R}^T C^{-1} \nu_{\mu R} \chi
+ \nu_{\tau R}^T C^{-1} \nu_{\tau R} \chi^\ast
\right) + {\rm H.c.}, \nonumber
\end{eqnarray}
and a $\nu_R$ mass term
\begin{eqnarray}
\lefteqn{\mathcal{L}_\mathrm{mass} =} \\ 
&& \frac{1}{2}\,
m^\ast \nu_{eR}^T C^{-1} \nu_{eR}
+ {m^\prime}^\ast \nu_{\mu R}^T C^{-1} \nu_{\tau R}
+ {\rm H.c.} \nonumber
\end{eqnarray}
We assume the VEV of $\chi$ and $m$, $m^\prime$ 
to be of the order of the seesaw scale.

\begin{table*}
\tbl{The four scalar multiplets with respect to the family symmetries 
  of the models of Sec.~\ref{class}. 
\label{scalars}}
{\begin{tabular}{@{}ccccccc@{}}
\toprule
Case & $\chi$ & $L_e$ & $L_\mu$ & $L_\tau$ & $\mu$--$\tau$ antisymm. 
& $\left( \mnu \right)^{-1}$ \\
\colrule
(1) & $\chi_{ee}$ & $-2$ & 0 & 0 & 
$\chi_{ee} \to -\chi_{ee}$ & $(ee) \neq 0$ \\  \colrule
(2) & $\chi_{\mu\tau}$ & 0 & $-1$ & $-1$ & 
$\chi_{\mu\tau} \to -\chi_{\mu\tau}$ & $(\mu\tau) \neq 0$ \\ \colrule
& $\chi_{e\mu}$ & $-1$ & $-1$ & 0 & & \\[-1.5mm] 
\raisebox{1.5ex}{(3)} 
& $\chi_{e\tau}$ & $-1$ & 0 & $-1$ & 
\raisebox{1.5ex}{$\chi_{e\mu} \leftrightarrow -\chi_{e\tau}$} &
\raisebox{1.5ex}{$(e\mu) \neq -(e\tau)$} \\ \colrule
& $\chi_{\mu\mu}$ & 0 & $-2$ & 0 & & \\[-1.5mm] 
\raisebox{1.5ex}{(4)} 
& $\chi_{\tau\tau}$ & 0 & 0 & $-2$ & 
\raisebox{1.5ex}{$\chi_{\mu\mu} \leftrightarrow -\chi_{\tau\tau}$} &
\raisebox{1.5ex}{$(\mu\mu) \neq -(\tau\tau)$} \\ \botrule
\end{tabular}}
\end{table*}
This model yields the inverse neutrino mass matrix\cite{GL05}
\begin{equation}\label{mnus3}
\left( \mnu \right)^{-1} = 
\left( \begin{array}{ccc} x & y & y \\ y & \hphantom{i} u\, e^{i\psi} & w \\
y & w & u\, e^{-i\psi} \end{array} \right).
\end{equation}
With the decomposition~(\ref{decomposition}) and Eqs.~(\ref{S}) 
and (\ref{AS}),
we find $z = u \cos \psi$, $q = i\,u \sin \psi$, $p = 0$.
For $\psi = 0$ or $\pi$, 
Eq.~(\ref{mnus3}) yields a $\mu$--$\tau$-symmetric neutrino
mass matrix. If $\psi \neq 0$, $\mu$--$\tau$ symmetry is partially broken,
such that the 
matrix of absolute values in $\left( \mnu \right)^{-1}$ is still
$\mu$--$\tau$-symmetric.
Which case is realized, depends on the type of symmetry breaking of $S_3$: If
it is broken spontaneously, then $\psi = 0$ or $\pi$; 
if, in addition, 
its $\mathbbm{Z}_3$ subgroup is broken softly via terms of dimension one and
two in the scalar potential, $\sin \psi$ is non-zero.
In the latter case, there are correlated deviations from Eq.~(\ref{a}),
approximately given by\cite{GL05}
\begin{eqnarray}
\lefteqn{\cos 2\theta_{23} \simeq -2\, c_{12} s_{12}} 
\\ && \times
\frac{\Delta m^2_\odot}{c_{12}^2 m_1^2 + s_{12}^2 m_2^2 - m_1^2 m_2^2/m_3^2}
\, s_{13} \cos \delta, \nonumber 
\end{eqnarray}
where $\dsol = m_2^2 - m_1^2$ is the solar mass-squared difference and
$\delta$ is the CKM-type phase in $U$.
For an inverted neutrino mass spectrum, $\theta_{23}$ is still maximal for all
practical purposes. For a normal spectrum, possible 
deviations of $\theta_{23}$ from
$45^\circ$ are most pronounced in the hierarchical case, 
namely $\cos 2\theta_{23} \sim -3\,s_{13} \cos \delta$.

\section{A class of models based on $\mu$--$\tau$ antisymmetry}
\label{class}
Here we discuss a class of models based on 
conserved lepton numbers and 
$\mu$--$\tau$ antisymmetry---see Sec.~\ref{framework}, Items~(\ref{fs}) and
(\ref{mu-tau}). Since a $\mu$--$\tau$-antisymmetric $M_R$ is singular,
we add complex scalar gauge singlets which carry lepton numbers. Such
scalars have the general Yukawa couplings
\begin{equation}\label{Lchi}
\mathcal{L}_Y(\chi) = \frac{1}{2} \sum_{\alpha,\beta} z_{\alpha\beta}^*\,
\nu_{\alpha R}^T C^{-1} \nu_{\beta R}\, \chi_{\alpha\beta}
+ \mbox{H.c.}
\end{equation}
In Table~\ref{scalars} we have listed the four basic cases 
of scalar singlets compatible with the family
symmetries. Their VEVs make $M_R$ non-singular and induce a 
$\mu$--$\tau$-symmetric contribution in $\left( \mnu
\right)^{-1}$---cf.\ Eq.~(\ref{decomposition})---as 
shown in the last column of Table~\ref{scalars}. 

Combining $M^{(\mathrm{AS})}$ with one or two of the cases in
Table~\ref{scalars} for the construction of $M_R$ leads to 
ten models---see Table~\ref{models}. Of these models, only five are
compatible with the data, as indicated in this table. Each of the five viable
models has six physical parameters in $\mnu$. Models (1)--(4) 
(four parameters in $\mnu$) and model~(10) (five parameters in $\mnu$)
are ruled out; properties of these models which lead to
contradiction with the data are found in the last column of
Table~\ref{models}. For the viable models, the preferred or predicted 
neutrino mass spectrum is indicated in that column.
\begin{table}
\tbl{The models which can be constructed by
  using one or
  two of the scalar multiplets of Table~\ref{scalars}. A cross
  (tick) indicates that the model is ruled out (allowed) by the data. The
  numbers (1)--(4) refer to the cases in Table~\ref{scalars}. 
\label{models}}
{\begin{tabular}{@{}rccc@{}}\toprule
\multicolumn{2}{c}{Case} & Validity & Comment \\ \colrule
 (1)\,: & --      & \ding{56} & $s_{13}^2 > 0.1$ \\
 (2)\,: & --      & \ding{56} & $s_{13} \tan 2\theta_{12}> 1$ \\
 (3)\,: & --      & \ding{56} & $\dsol/\datm > 1$ \\
 (4)\,: & --      & \ding{56} & $\dsol/\datm > 1$ \\
 (5)\,: & (1)+(2) & \ding{52} & any spectrum \\
 (6)\,: & (1)+(3) & \ding{52} & normal preferred \\
 (7)\,: & (1)+(4) & \ding{52} & any spectrum\\
 (8)\,: & (2)+(3) & \ding{52} & inverted spec. \\
 (9)\,: & (2)+(4) & \ding{52} & inverted spec. \\
(10)\,: & (3)+(4) & \ding{56} & $\dsol/\datm > 1$ \\ 
\botrule
\end{tabular}}
\end{table}

Let us make some comments on the models of this section. 
Whenever 
$\left( \mnu^{-1} \right)_{ee} = 
\left( \mnu^{-1} \right)_{\mu\tau} = 0$, 
then $\dsol/\datm > 1$, where 
$\datm = \left| m_3^2 - m_1^2 \right|$ is the 
atmospheric mass-squared difference. This is the case for models
(3), (4), (10) and the reason why they are ruled out. If 
$\left( \mnu^{-1} \right)_{ee} = 0$, then only the inverted neutrino mass
spectrum is possible. Among the allowed models, this applies to (8)
and (9). Finally we want to mention that model~(8) is the most
predictive one; e.g., a slight deviation of 
$\sin^2 2\theta_{23}$ from one leads to a large $s_{13}^2$, which in
practice gives the lower bound 
$\sin^2 2\theta_{23} > 0.99$.
For further details see Ref.~\refcite{anti}.

\balance

\section{Conclusions}
In this report we have combined a 
$\mu$--$\tau$ interchange symmetry with family symmetries which give
diagonal Yukawa couplings in order to obtain a predictive
neutrino mass matrix. We have considered extensions of the SM which
have three Higgs doublets and right-handed neutrino singlets for the
seesaw mechanism. Other important ingredients are scalar gauge
singlets which induce, upon acquiring VEVs, contributions to $M_R$.
With diagonal Yukawa couplings, lepton mixing stems solely from a
non-diagonal $M_R$ and conditions on $M_R$ are translated into
conditions on $\left( \mnu \right)^{-1}$.
While 
exact $\mu$--$\tau$ symmetry in $\mnu$ or 
$\left( \mnu \right)^{-1}$ leads to Eq.~(\ref{a}), 
deviations from exact $\mu$--$\tau$ symmetry can lead to interesting
correlations between atmospheric mixing and $\theta_{13}$.
Though exact $\mu$--$\tau$ antisymmetry in $\mnu$ or 
$\left( \mnu \right)^{-1}$ is not viable, it is nevertheless a
useful concept,
in combination with the above-mentioned scalar singlets, for producing
predictive models.

\section*{Acknowledgments}
The author is grateful to L.\ Lavoura and B.\ Strohmaier 
for a critical reading of the manuscript.


\begin{thebibliography}{9}

\bibitem{tortola}
See, e.g., 
M. Maltoni, T. Schwetz, M. T\'ortola and J.W.F. Valle,
\Journal{New J. Phys.}{6}{122}{2004};
G.L. Fogli, E. Lisi, A. Marrone and A. Palazzo,
\Journal{Prog. Part. Nucl. Phys.}{57}{742}{2006}.

\bibitem{early}
T. Fukuyama and H. Nishiura, 
hep-ph/9702253, in Proc. of \textit{International Workshop on Masses
  and Mixings of Quarks and Leptons}, ed. Y. Koide 
(World Scientific, Singapore, 1998); 
R.N. Mohapatra and S. Nussinov,
\Journal{Phys. Rev.}{D60}{013002}{1999};
E. Ma and M. Raidal, 
\Journal{Phys. Rev. Lett.}{87}{011802}{2001} 
[\Journal{Err. ibid.}{87}{159901}{2001}];
C.S. Lam, 
\Journal{Phys. Lett.}{B507}{214}{2001};
K.R.S. Balaji, W. Grimus and T. Schwetz,
\Journal{Phys. Lett.}{B508}{301}{2001};
E. Ma,
\Journal{Phys. Rev.}{D66}{117301}{2002};
P.F. Harrison and W.G. Scott, 
\Journal{Phys. Lett.}{B547}{219}{2002}.

\bibitem{Z2model}
W. Grimus and L. Lavoura,
\Journal{JHEP}{07}{045}{2001};
W. Grimus and L. Lavoura,
\Journal{Acta Phys. Pol.}{B32}{3719}{2001}.

\bibitem{ustron}
W. Grimus and L. Lavoura, 
\Journal{Acta. Phys. Pol.}{B34}{5393}{2003}.

\bibitem{aizawa}
T. Kitabayashi and M. Yasu\`e,
\Journal{Phys. Lett.}{B621}{133}{2005};
I. Aizawa, T. Kitabayashi and M. Yasu\`e,
\Journal{Nucl. Phys.}{B728}{220}{2005};
T. Kitabayashi and M. Yasu\`e,
\Journal{Phys. Rev.}{D73}{015002}{2006}.

\bibitem{anti}
W. Grimus, S. Kaneko, L. Lavoura, H. Sawanaka and M. Tanimoto, 
\Journal{JHEP}{01}{110}{2006}.

\bibitem{lavoura}
L. Lavoura, 
\Journal{Phys. Lett.}{B609}{317}{2005}.

\bibitem{GL05}
W. Grimus and L. Lavoura, 
\Journal{JHEP}{08}{013}{2005}.

\end{thebibliography}
\end{document}